\documentclass[twocolumn, astrosymb, tighten]{aastex631}
\usepackage{subfigure} 

\shorttitle{The Argus Optical Array}
\shortauthors{Law et al.}

\graphicspath{{./}{figures/}}

\begin{document}

\title{Low-Cost Access to the Deep, High-Cadence Sky: the Argus Optical Array}

\correspondingauthor{Nicholas Law \& Hank Corbett} \email{nmlaw@physics.unc.edu, htc@unc.edu}

\author{Nicholas M. Law}
\affiliation{Department of Physics and Astronomy, University of North Carolina at Chapel Hill, Chapel Hill, NC 27599-3255, USA}

\author{Hank Corbett}
\affiliation{Department of Physics and Astronomy, University of North Carolina at Chapel Hill, Chapel Hill, NC 27599-3255, USA}

\author{Nathan W. Galliher}
\affiliation{Department of Physics and Astronomy, University of North Carolina at Chapel Hill, Chapel Hill, NC 27599-3255, USA}

\author{Ramses Gonzalez}
\affiliation{Department of Physics and Astronomy, University of North Carolina at Chapel Hill, Chapel Hill, NC 27599-3255, USA}

\author{Alan Vasquez}
\affiliation{Department of Physics and Astronomy, University of North Carolina at Chapel Hill, Chapel Hill, NC 27599-3255, USA}

\author{Glenn Walters}
\affiliation{Department of Applied Physical Sciences, University of North Carolina at Chapel Hill, Chapel Hill, NC 27599-3050, USA}

\author{Lawrence Machia}
\affiliation{Department of Physics and Astronomy, University of North Carolina at Chapel Hill, Chapel Hill, NC 27599-3255, USA}

\author{Jeff Ratzloff}
\affiliation{Department of Physics and Astronomy, University of North Carolina at Chapel Hill, Chapel Hill, NC 27599-3255, USA}
\affiliation{Corvid Technologies, 153 Langtree Campus Drive, Mooresville, NC 28117}

\author{Kendall Ackley}
\affiliation{Department of Physics, University of Warwick, Gibbet Hill Road, Coventry CV4 7AL, UK}

\author{Chris Bizon}
\affiliation{Renaissance Computing Institute, 100 Europa Drive Suite 540,
Chapel Hill, North Carolina 27517}

\author{Christopher Clemens}
\affiliation{Department of Physics and Astronomy, University of North Carolina at Chapel Hill, Chapel Hill, NC 27599-3255, USA}

\author{Steven Cox}
\affiliation{Renaissance Computing Institute, 100 Europa Drive Suite 540,
Chapel Hill, North Carolina 27517}

\author{Steven Eikenberry}
\affiliation{Department of Astronomy, 211 Bryant Space Science Center, Gainesville, FL 32611-2055, USA}

\author{Ward S. Howard}
\affiliation{Department of Physics and Astronomy, University of North Carolina at Chapel Hill, Chapel Hill, NC 27599-3255, USA}

\author{Amy Glazier}
\affiliation{Department of Physics and Astronomy, University of North Carolina at Chapel Hill, Chapel Hill, NC 27599-3255, USA}

\author{Andrew W. Mann}
\affiliation{Department of Physics and Astronomy, University of North Carolina at Chapel Hill, Chapel Hill, NC 27599-3255, USA}

\author{Robert Quimby}
\affiliation{San Diego State University, 5500 Campanile Dr., San Diego, CA 92182, USA}

\author{Daniel Reichart}
\affiliation{Department of Physics and Astronomy, University of North Carolina at Chapel Hill, Chapel Hill, NC 27599-3255, USA}

\author{David Trilling}
\affiliation{Department of Astronomy and Planetary Science, Northern Arizona University, 527 S. Beaver St., Flagstaff, Arizona 86011-6010}

\begin{abstract}
New mass-produced, wide-field, small-aperture telescopes have the potential to revolutionize ground-based astronomy by greatly reducing the cost of collecting area. In this paper, we introduce a new class of large telescope based on these advances: an all-sky, arcsecond-resolution, 1000-telescope array which builds a simultaneously high-cadence and deep survey by observing the entire sky all night. As a concrete example, we describe the Argus Array, a 5m-class telescope with an all-sky field of view and the ability to reach extremely high cadences using low-noise CMOS detectors. Each 55 GPix Argus exposure covers 20\% of the entire sky to $\rm{m_g}$=19.6 each minute and $\rm{m_g}$=21.9 each hour; a high-speed mode will allow sub-second survey cadences for short times. Deep coadds will reach $\rm{m_g}$=23.6 every five nights over 47\% of the sky; a larger-aperture array telescope, with an étendue close to the Rubin Observatory, could reach $\rm{m_g}$=24.3 in five nights. These arrays can build two-color, million-epoch movies of the sky, enabling sensitive and rapid searches for high-speed transients, fast-radio-burst counterparts, gravitational-wave counterparts, exoplanet microlensing events, occultations by distant solar system bodies, and myriad other phenomena. An array of O(1,000) telescopes, however, would be one of the most complex astronomical instruments yet built. Standard arrays with hundreds of tracking mounts entail thousands of moving parts and exposed optics, and maintenance costs would rapidly outpace the mass-produced-hardware cost savings compared to a monolithic large telescope. We discuss how to greatly reduce operations costs by placing all optics in a thermally controlled, sealed dome with a single moving part. Coupled with careful software scope control and use of existing pipelines, we show that the Argus Array could become the deepest and fastest Northern sky survey, with total costs below \$20M.
\end{abstract}

\section{Introduction} \label{sec:intro}
Almost all time-domain surveys tile across the sky, achieving depth with large collecting areas, and cadence by returning to previously-visited patches after some time. This approach has been enormously successful, with a large number of surveys including, for example, the Palomar Transient Factory  \citep[PTF;][]{Rau2009}, the Mobile Astronomical System of Telescope-Robots \citep[MASTER;][]{master_instrument},  the Asteroid Terrestrial-impact Last Alert System \citep[ATLAS;][]{atlas_instrument}, the All-Sky Automated Survey for Supernovae \citep[ASAS-SN;][]{asassn_instrument}, the Zwicky Transient Facility \citep[ZTF;][]{ztf_instrument}, Pan-STARRS \citep{panstarrs_instrument}, the Catalina Sky Survey \citep[CSS;][]{catalina_instrument} and Catalina Real-Time Transient Survey \citep[CRTS;][]{crts_instrument}, the Dark Energy Survey \citep[DES;][]{des_program}, and the Gravitational Wave Optical Transient Observatory \citep[GOTO;][]{goto_instrument}. Covering the sky piece-by-piece with a tiling survey, however, returns to each part of the sky only on much longer timescales than the survey's exposure time. 

Short-timescale astronomical phenomena such as optical fast-radio-burst counterparts, kilonovae, small-planet microlensing, and a host of others, are thus necessarily relegated to deep-drilling fields where the survey field is a small multiple of the field of view of the telescope. Even so, the new development of scientific-grade CMOS devices has enabled a host of new surveys using second-timescale (or faster) cadences to reach these phenomena, including the Organized Autotelescopes for Serendipitous Events Survey \citep[][OASES;]{arimatsu_2017}, the Transneptunian Automated Occultation Survey \citep[][TAOS-II;]{wang_2016}, the Colibri photometry array \citep{pass_2018}, the Tomo-e Gozen instrument on the Kiso Schmidt telescope \citep{sako_2018, richmond_2019}, the prototype drift scan system presented in \citep{tingay_2020, tingay_2021}, and the Weizmann Fast Astronomical Survey Telescope \citep{Nir_2021}.

In this paper, we discuss a concept for a new type of high-cadence time-domain survey: a large telescope with a distributed aperture that observes the entire sky simultaneously. Using an array of mass-produced small telescopes and high-speed detectors, the telescope can simultaneously achieve high cadence and sky coverage, and produce a deep survey by long-term coaddition of images. Instead of minute-timescale tiles of the deep sky every few days, the array will produce continuous nighttime coverage of every part of the sky, along with deep sky observations on longer, coadded, timescales. As a consequence of pushing to a field of view of the entire sky, the concept of pointing the telescope is retired: the array will always be pointing at phenomena as they happen, everywhere, generating a deep movie of the sky before, during, and after events of all types.

This concept is enabled by a revolution of capabilities in the amateur astronomy community. In 2017, Celestron introduced the Rowe-Ackermann Schmidt Astrograph (RASA) line of telescopes \citep{rasa_whitepaper}, which match good image quality across a wide field with low-cost mass production. List prices are only around \$60,000 per $\rm{m^2}$ of glass, including all secondary optics and ancillary mounting components. Especially when coupled with wide-field sensors, such as the new tens-of-MPix ultra-low-noise CMOS detectors from Sony, arrays of these telescopes offer an affordable route to building large scale telescopes. Several groups are exploring using this hardware to build several-meter-aperture-class telescopes which are capable of repointing their telescopes to rapidly trade off field of view with effective aperture -- for example the LAST telescope \citep{ofek_2020}. 

\begin{figure}
	\centering
		\includegraphics[width=\columnwidth]{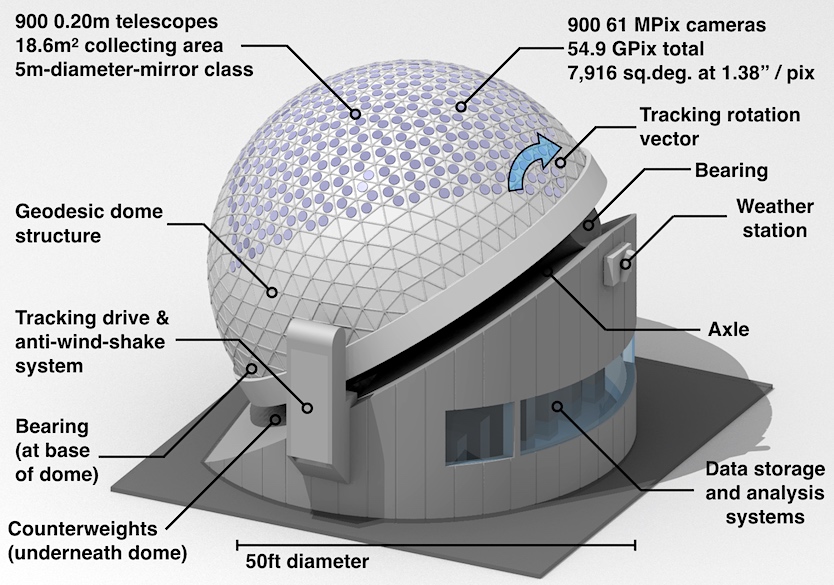}
	\caption{An overview of the Argus Array concept: a thermally-controlled, weather-sealed hemisphere mounting hundreds of telescopes on a single tracking mount. The stability of the telescope environment and the single moving part used during the night greatly simplifies operations and maintenance compared to a large array of individual telescopes.}
    \label{fig:argus_concept}
\end{figure}    

The move to building ever larger array telescopes, however, comes with new complexities. In a conventional design, with a field of telescopes on individual or grouped mounts, each telescope is subjected to day/night thermal cycles and the resultant effects on optical alignment and camera integrity, as well as dust and dirt accumulation on optics and moving parts.  Hundreds of mounts must be kept operational and tracking at arcsecond-level performance, and each telescope requires at least a focuser, tube cooling and camera cooling. In total, a large telescope array will have thousands of moving parts and at least many hundreds of exposed optical surfaces to keep clean and aligned. 

The prosaic problems of night-to-night telescope operation -- keeping the optics clean, maintaining image quality, and performing standard moving-part maintenance -- thus have the potential to become an overwhelming challenge: a single, simple maintenance procedure requiring a few hours of work on each telescope once per year scales to requiring multiple full-time staff members, just for that single procedure. Furthermore, given the resulting year-long continuous maintenance cycles, this precludes operating with a stable array: telescopes will need to be continually swapped in and out. This constantly-changing array will greatly complicate the task of building image calibrations, pipeline data analysis, and ultimately producing consistent results.

The Argus Optical Array concept we discuss in this paper as a concrete example of this class of telescopes (Figure \ref{fig:argus_concept}) is designed to greatly reduce these maintenance and operations problems, while observing the entire available sky simultaneously. It combines the low-cost wide-field capabilities of a large mass-produced telescope array with the relatively low operations and maintenance costs of a conventional monolithic telescope. Compared to a conventional telescope array, the Argus concept loses flexibility in survey design, being unable to individually point telescopes and thus precluding combining telescope apertures to form a single large telescope. Instead, it covers the entire accessible sky all night, and achieves depth by long integrations on each part of the sky. By sealing all optical surfaces within a thermally controlled environment, mounting all telescopes on a single tracking mount, and using the telescope structure itself to protect the optics from the weather, the Argus Optical Array concept can open up the possibility of building extremely large telescope arrays at overall costs far lower than those of equivalent conventional telescopes.

\begin{figure*}
	\centering
		\includegraphics[width=\textwidth]{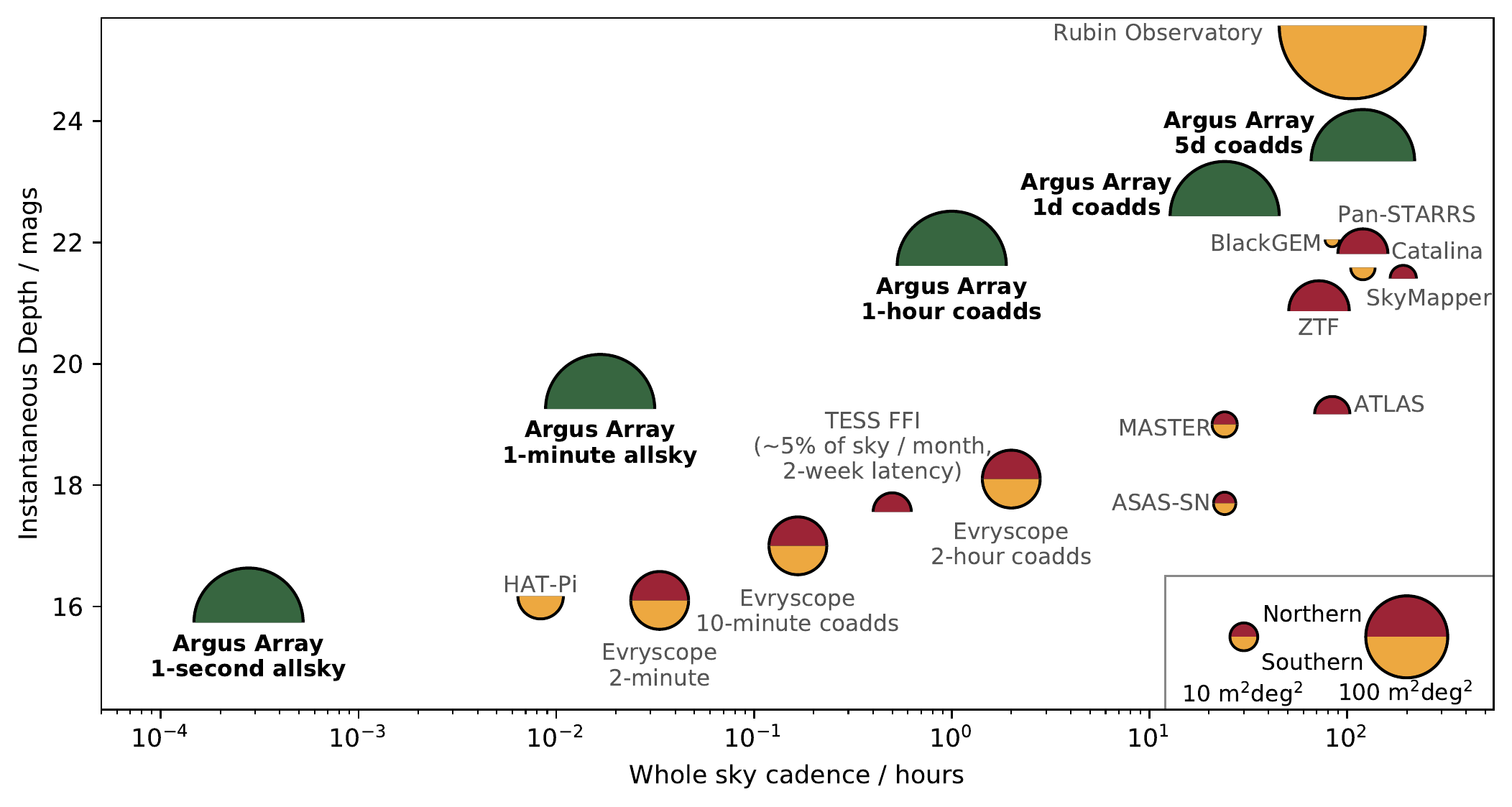}
		
	\caption{Operating and planned all-sky surveys with rapid transient detection capabilities. The position of each survey is a function of mirror, étendue, cadence and survey design. The shape and size of the points shows site location and étendue respectively. Where telescopes operate multiple surveys, we have selected the survey with the fastest rate of all-sky coverage, and (where available in the literature) we compare dark-sky, 5$\sigma$-detection limiting magnitudes. The Argus concept builds depth by all-night coadding of short exposures on each part of the sky, producing a hybrid fast and deep survey. The array's predicted performance is based on on-sky measurements, individual exposures and coadded, from individual telescopes, and an exposure-time calculator validated with existing sky surveys. The performance shown here is for the 8-inch-aperture Argus design; the 11-inch design (see appendix) affords approximately 0.7-magnitudes deeper performance, at increased cost.}
    \label{fig:survey_comp}
\end{figure*}        

We discuss the survey performance of an all-sky arcsecond-resolution array, and its key hardware parameters, in section \ref{sec:perf}. We describe the key science cases enabled in Section \ref{sec:science}. In Section \ref{sec:design} we discuss how to actually build a large telescope array, including maintaining the large number of telescopes, and the challenges of building software pipelines to analyze the resulting very high data rate. We describe the Argus Array Prototype systems currently under construction in section \ref{sec:conclusions}.

\section{The performance of currently-feasible all-sky large telescopes}
\label{sec:perf}
In this section we delineate the currently-achievable performance of a large, arcsecond-resolution, all-sky telescope array. We take as a goal the coverage of the entire accessible sky, down to an airmass limit, with a sufficiently large field of view so that every part of the sky gets many hours of continuous coverage through the night. We assume the telescopes in the array are arranged to tile across the sky without gaps, and that the array tracks the sky for some period of time, followed by a reset and starting tracking again from a new position (a ``ratchet'').

\subsection{Key hardware parameters}
For the all-sky telescope concept the field of view must be sufficient to enable deep coadding through the night without repointing. Practically, this requires at least an 8,000 square degree field of view \citep{law-evryscope-2015}. The Evryscopes \citep{ratzloff-evryscope} have pioneered the all-sky-telescope approach, covering the sky with a 1.4 GPix array capable of monitoring events brighter than $\rm m_g=16$. However, despite an \'etendue of $\rm 48 m^2 deg^2$, larger than almost all other sky surveys, the Evryscopes cannot begin to match the actual performance of large-telescope sky surveys: their 13\arcsec pixels mean the systems must work with per-pixel sky brightnesses hundreds of times greater than a typical survey. Building a truly deep simultaneous-all-sky survey requires, therefore, achieving arcsecond-scale pixels across the sky. One-arcsecond sampling over 8,000 square degrees requires O($10^{11}$) pixels. For the 50-100 MPix detectors currently available at reasonable prices this leads to a requirement of operating on the order of 1,000 telescopes.

\begin{figure*}
     \centering
     \begin{subfigure}
         \centering
         \includegraphics[width=0.48\textwidth]{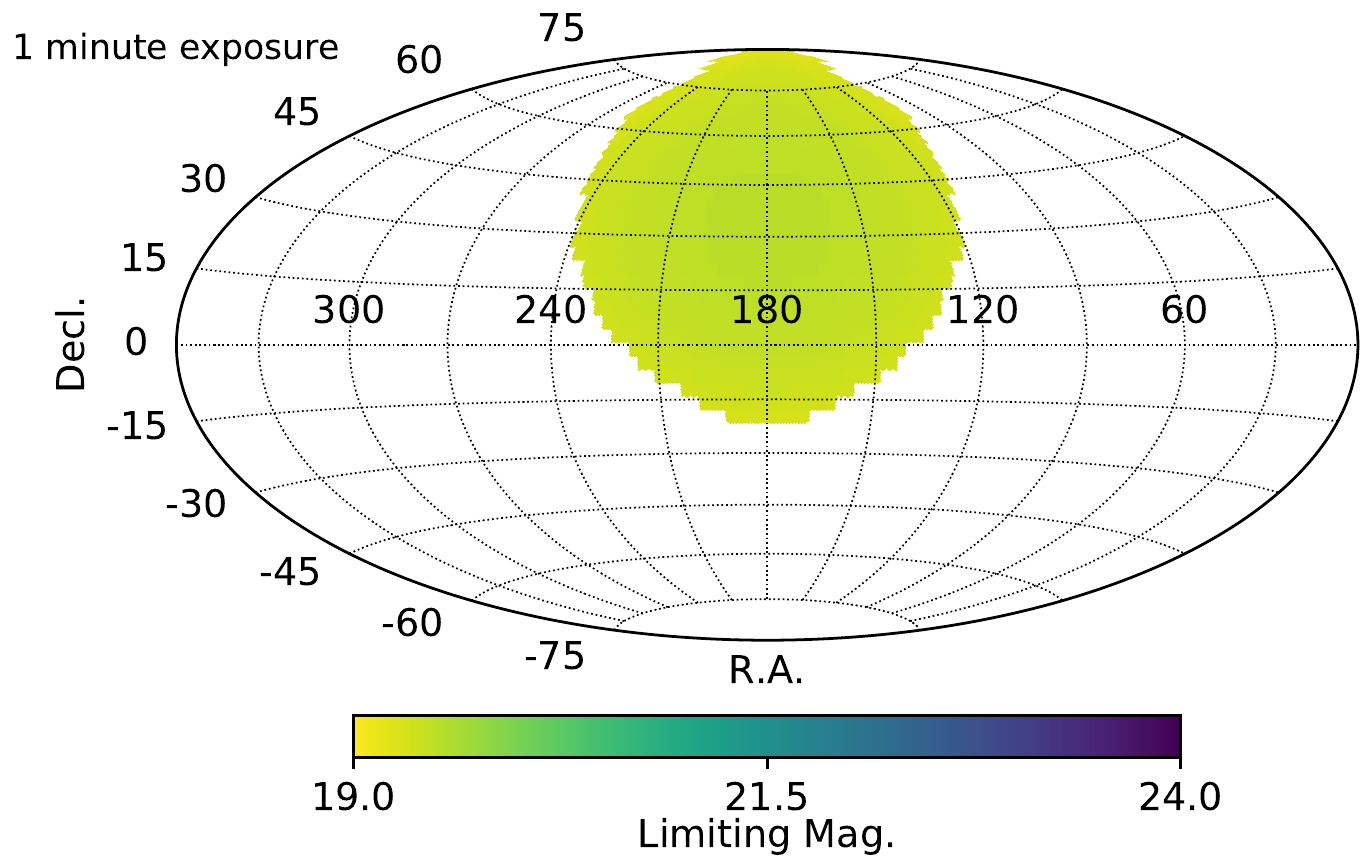}
     \end{subfigure}
     \hfill
     \begin{subfigure}
         \centering
         \includegraphics[width=0.48\textwidth]{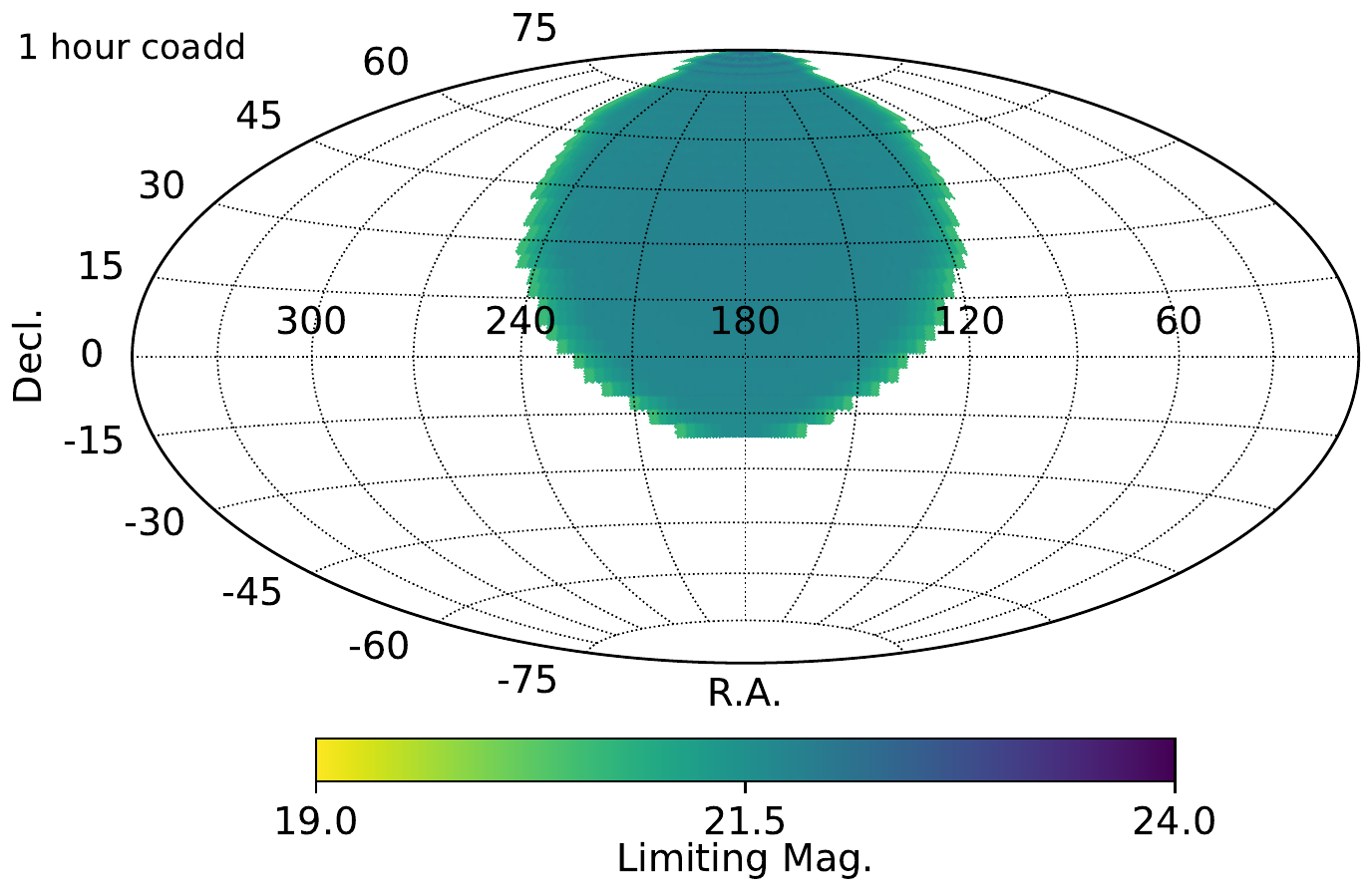}
     \end{subfigure}

     \begin{subfigure}
         \centering
         \includegraphics[width=0.48\textwidth]{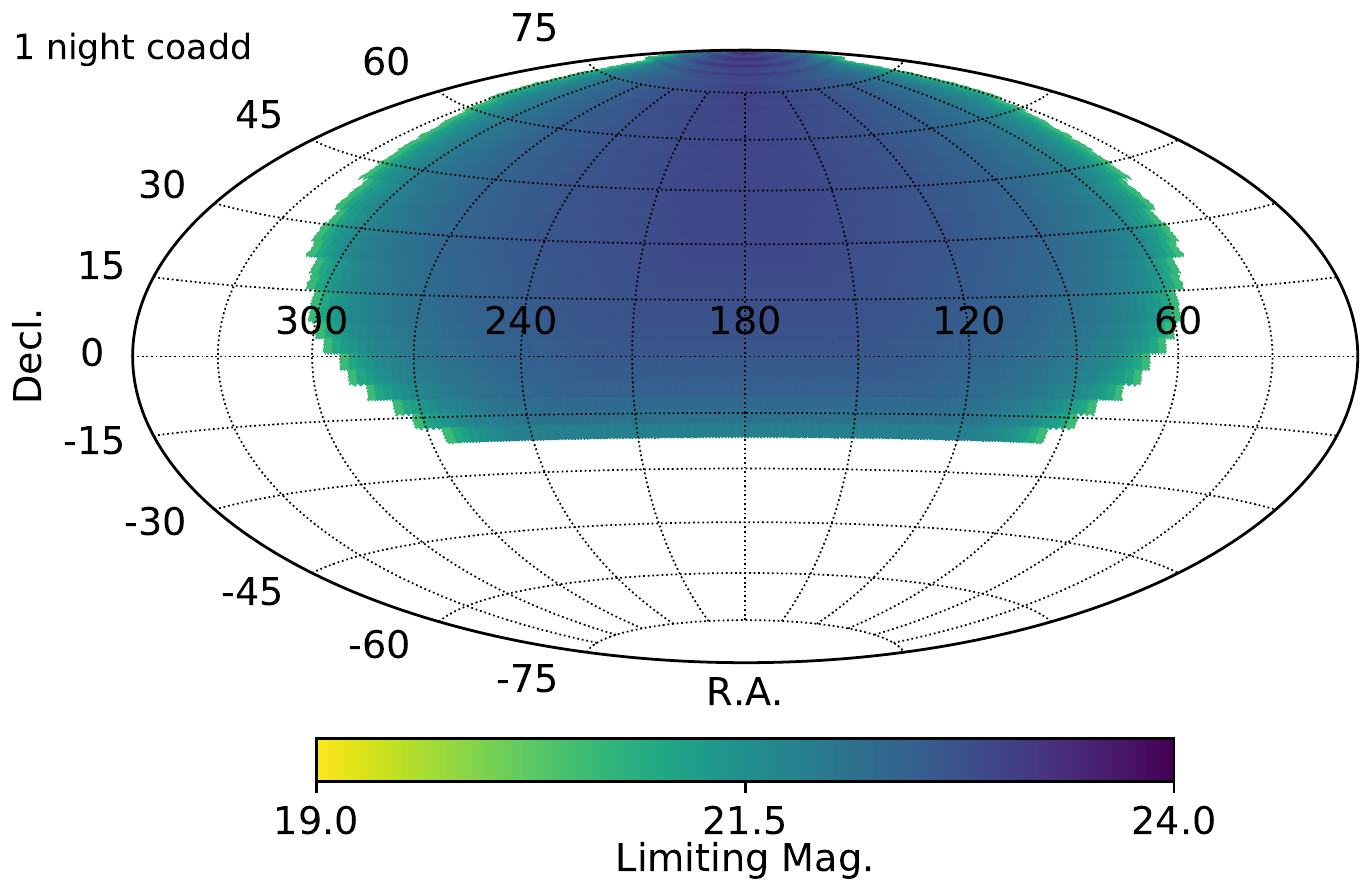}
     \end{subfigure}
     \hfill
     \begin{subfigure}
         \centering
         \includegraphics[width=0.48\textwidth]{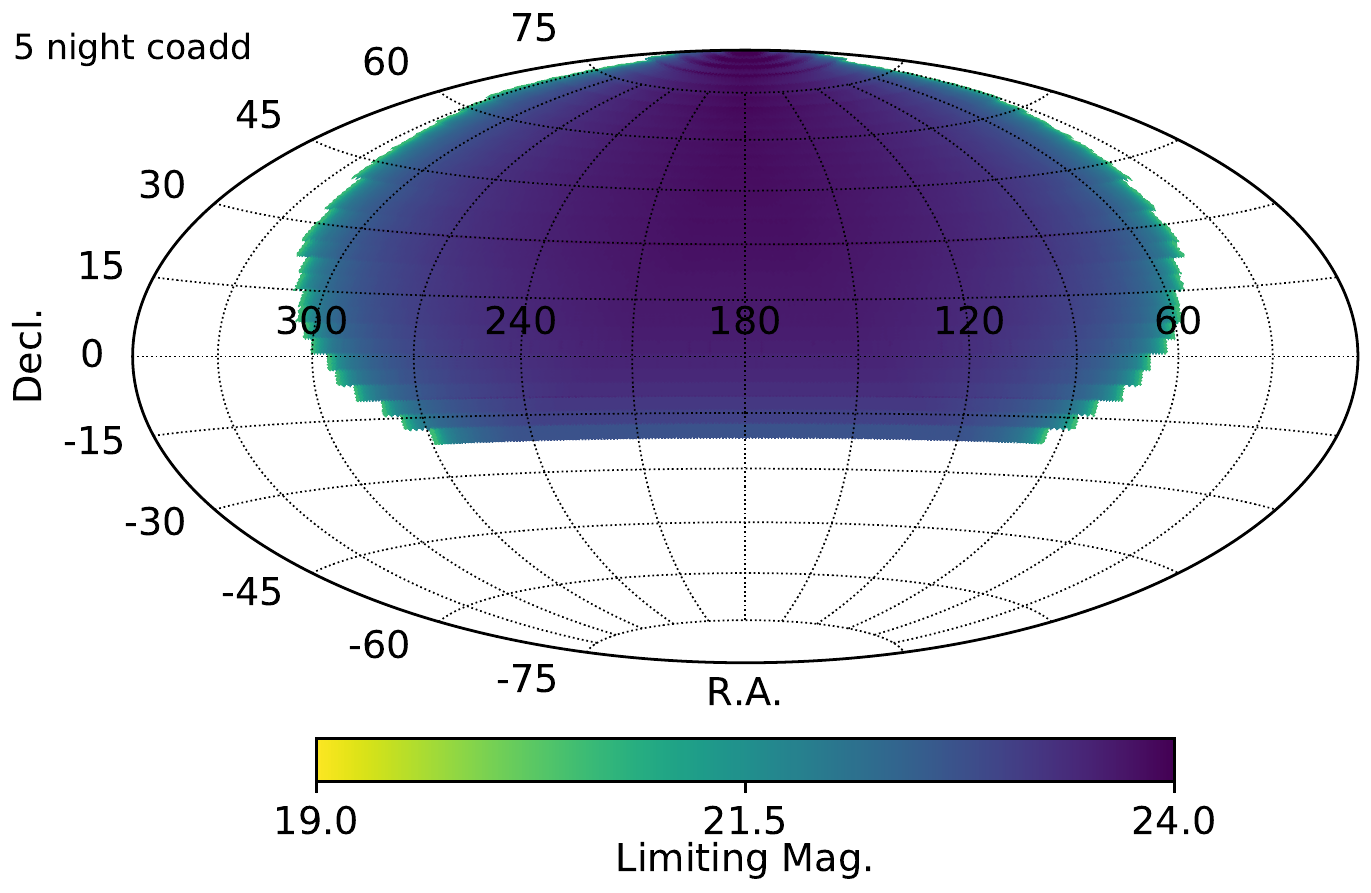}
     \end{subfigure}

    \caption{The limiting magnitude and sky coverage of the Argus Optical Array building up with time (60 seconds, 1 hour, 1 day, 5 days). Limiting magnitudes are for 5$\sigma$ signal to noise ratio, for the hardware described in section \ref{sec:perf}, and Monte-Carlo simulated for median conditions including a model of moon brightness and weather at a site similar to the Argus deployment site.}
    \label{fig:sky_coverage}
\end{figure*}

Current commercial wide-field, mass-produced telescopes are available at large-array-feasible prices with apertures between 20 and 28 cm. Because of the simple survey design, just the combination of field of view, sampling and aperture is sufficient to explore the science performance of a deep-all-sky telescope. For the following discussion we set the parameter space as follows: 1) 8,000 square degree field of view; 2) commercially-available mass-produced telescopes, and 3) using the largest commodity-priced sensors to achieve arcsecond-scale sampling across the sky. 

To avoid confusion and excessive complication with a variety of hardware choices, we focus the rest of the paper on a baseline 20-cm-telescope design. Although the RASA-8 telescope available from Celestron would be an excellent choice for a large array telescope, we are currently basing our designs on a mass-produced telescope custom-made for the Argus Optical Array by Planewave Instruments, that offers a Cassegrain-focus imaging position and improved sky sampling. The appendix details the specific hardware choices for the Argus Optical Array, and for a concrete discussion of the science possibilities from a real all-sky telescope design, in Table \ref{tab:argus_specs} we detail the specifications of the Argus Array. 

An array based on the RASA 11-inch telescope would produce an $\sim$0.6-magnitude deeper survey than the Argus-8 survey. A RASA-11 array would be significantly more expensive, because it requires more telescopes and cameras and the telescopes themselves are 3$\times$ more expensive. The telescopes are also more than 2$\times$ larger and heavier, and the array structure to hold that many telescopes will therefore be significantly more expensive than the 8-inch design. The array telescope design itself, however, scales well to that telescope size, and the RASA-11 remains a strong contender for a larger and more capable survey. Like all telescopes, the ultimate choice of aperture will depend on the funding available. We also detail the performance of an upgraded 11-inch-aperture array in the text, where appropriate.

\subsection{Survey design: cadence and wavelength coverage}
With no ability (or need) to point the individual telescopes, the only choices of survey design for an all-sky telescope array are the imaging cadence and the observing wavelengths of the telescopes. 

In cadence, newly-available CMOS detectors enable extremely short exposure times due to their low read noise and high sensitivity (see the appendix), reaching sub-second cadence with high duty cycles. The resulting data rates, however, would be extremely complex and expensive to deal with in realtime at full resolution, and the images would be read noise limited under most conditions, precluding deep coadding. For that reason, we suggest an all-sky array should have two modes of operation: 1) a standard minute-cadence survey mode where the incoming images are analyzed in realtime, full-cadence stamps are made around new sources and targets selected by the science team, and full imaging data is coadded and stored on single-ratchet timescales; and 2) a high-speed burst mode to be used in bright time, with integration times as short as 0.25s, where the system performs simple realtime image-to-image differencing and stores full-speed cutouts around regions of interest (all detectable stars \& transient-host galaxies, for example)\footnote{an ROI mode implemented on the sensor itself, potentially available for the Argus cameras, would allow far faster cadences in the hundreds of frames per second range.}

In wavelength, the advantages of multi-color observing likely outweigh the cost of increased complexity (see section \ref{sec:science}). The ratcheting all-sky survey design lends itself to alternating coverage, where telescopes adjacent in RA have different filters. One possible design for the coverage includes a g-band filter \citep[][]{1996AJ....111.1748F} and a wider g+r filter. Each part of the sky therefore receives 9 minutes of narrow-band coverage followed by 9 minutes of wide-band coverage. For transient sources that evolve more slowly than minute timescales, subtracting adjacent ratchets (or modelling the time evolution) would provide separate g and r lightcurves, while receiving 50\% more light over the survey lifetime than using two narrow bands.
     
\subsection{Survey Depth}
Compared to current and planned all-sky surveys (Figure \ref{fig:survey_comp}), an all-sky 20-cm-telescope array could become the deepest Northern-hemisphere time-domain sky survey, as expected given its 5m-class total collecting area. The array's \'etendue would be half that of the Rubin Observatory, despite the array's much-lower hardware costs of $<$\$10M. Beyond a simple \'etendue comparison (which does not capture many relevant parameters \citep[][]{Tonry_2011, law-evryscope-2015, ofek_2020}, the array's small pixels compared to surveys like the Evryscopes are what enable it to become competitive with the largest current tiling-survey telescopes, with a depth midway between the largest current Northern-hemisphere sky surveys and the Rubin Observatory. This is possible because, although each individual telescope has less than one-thousandth of the collecting area of the Rubin Observatory's mirror, their combined field of view means that they can observe each part of the sky for more than 2,000 times longer over the survey lifetime.

\subsubsection{Survey depth calculations}
We calculate the array's limiting magnitudes taking into account the hardware optical quality, aperture photometry with optimally-sized apertures, optics and atmospheric throughput, measured quantum efficiencies for the detectors, the airmass of each individual telescope. Integrating over wavelength, we calculate the filter, quantum efficiency, and atmospheric throughput (using MODTRAN, \citealt{Berk_2014}) effects, and assume 70\% telescope throughput. Specifically for the Argus Optical Array, we use an aperture diameter of 3 pixels (matched to the Argus-8's stable PSFs across the field of view), and estimate that, based on on-sky measurements and optical simulations, this will capture 80\% of the source photons that make it to the sensor. When coadding images, we have calculated scalings based on theoretical noise calculations and on-sky measurements with real hardware.

\begin{deluxetable}{ll}
\tablecaption{\label{tab:argus_specs}The specifications of the Argus Optical Array}
\tabletypesize{\footnotesize}
\startdata
\vspace{0.25cm}\\
\bf Hardware & \\
\hline
\noalign{\vskip 1mm}   
System design & 900 telescopes; shared tracking drive\\
              & Sealed dome for long-term stability\\
Telescopes & 203 mm aperture, F/2.8\\
Detectors & 62 MPix Sony IMX455 CMOS sensor\\
& 1.7e- noise @ 0.4s full-frame readout\\
& $\approx$90\% QE @ 500nm\\
&$2\times10^4$ e- well capacity \\
Total detector size & 54.9 GPix\\
Field of view & 9.04 sq. deg. per telescope\\
              & 7916 sq.deg. instantaneous total\\
Sky coverage per night & 19,370 sq. deg. (2-10 hours per night)\\
Sampling & 1.38\arcsec/pixel\\
Site         & North America\\
Exposure times & 1 sec \& 1 minute\\
Observing strategy & Track for 9 minutes;\\ 
                   & reset (48s) and repeat ("ratchet")\\
Wavelengths & 9 minutes (1 ratchet) in g-band\\
                   & followed by 9 minutes in wide-band\\
Data analysis & Realtime analysis for fast transients \\
              & Full-data storage at 15-minute cadence\\
\vspace{0.25cm}\\
\bf Limiting magnitudes & 5$\sigma$, median sky conditions \\
\hline
\noalign{\vskip 1mm}   
High-speed & $\rm{m_g}$=16.1 in 1 sec.; 20\% of entire sky\\
Normal operation & $\rm{m_g}$=19.6 in 1 min.; 20\% of entire sky\\
\vspace{0.1cm}
Coadding & $\rm{m_g}$=21.9 in 1 hour; 24\% of entire sky\\
         & $\rm{m_g}$=22.7 in 1 nt.; 47\% of entire sky\\
         & $\rm{m_g}$=23.6 in 5 nts.; 48\% of entire sky\\
\enddata
\end{deluxetable}

Given the all-sky array's size and extremely-wide sky coverage, it is not sufficient to simply estimate the limiting magnitude for a single telescope or pointing to obtain the full survey characteristics. We constructed a Monte-Carlo simulation to sample the limiting magnitudes across the entire array at a variety of different weather and moon conditions. In each iteration of the simulation, we sample a random telescope position taken from one of the hundreds of telescopes in the array and use the corresponding airmass to scale the loss of atmospheric transmission. For dark-sky observations we assume a g-band sky brightness of 22.1, a canonical value based on measured values at Apache Point \citep{Stoughton2002, Abazajian_2003}, Haleakala \citep{Chambers_2016} and Cerro Pachon \citep{Ivezi_2019}.  For median sky conditions numbers we sample a sky brightness from the all-year, multiple-sky-position distribution of measurements taken by the Sloan Digital Sky Survey at Apache Point \citep{Stoughton2002, Abazajian_2003}. We sample 50,000 combinations of weather and array telescope positions to allow an estimation of the median limiting magnitude over the various sky conditions and array pointings. We simulate the array coadding images with realistic ratchet timescales and array arrangements, measuring the number of images taken in each part of the sky as the array builds up coverage through hours and nights (Figure \ref{fig:sky_coverage}).

In each 8,000 square degree one-minute exposure, an all-sky array based on the Argus-8 telescope\footnote{the 11-inch-aperture upgraded array could reach $\rm{m_g}$=20.3 per minute, $\rm{m_g}$=22.6 per hour, $\rm{m_g}$=23.4 per night, and $\rm{m_g}$=24.3 in five nights.} will reach a limiting magnitude of 19.6 in g-band (Figure \ref{fig:sky_coverage}; 5$\sigma$ detections , median sky conditions). Coadding will increase this to $\rm{m_g}$=21.9 over 1 hour, $\rm{m_g}$=22.7 every night over 47\% of the sky, and $\rm{m_g}$=23.6 every five nights. The dark-sky limiting magnitude of the system, including all airmass effects from the array size, is approximately 0.15 magnitudes deeper, at  $\rm{m_g}$=19.7, $\rm{m_g}$=22.0, $\rm{m_g}$=22.9, and $\rm{m_g}$=23.7 respectively.

The array's longer image coaddition depths are not quite as deep as might be expected from a simple calculation of the number of images per night; this is because over one night the array's pointing positions move with the meridian, so each part of the sky gets about half a night of observation (more towards the poles). The longer image coaddition depths are averages across the whole array, and account for varying observability windows as a function of declination. Given the $110^{\circ}$ declination range covered, the array observes approximately half of the entire sky (19,370 square degrees) each night. 

Based on Gaia star counts \citep[][]{GAIA-mission, GAIA-dr2} and TRILEGAL simulations for fainter stars \citep{trilegal}, we estimate that each array exposure will include $10^8$ stars (somewhat depending on season). Each night, with coadds going deeper and the increased sky area from sweeping across the available sky during the night, the Array will observe around $10^9$ stars.

In high-speed burst mode the Array will be able to reach magnitudes of $\rm m_g\approx16.1$ in one second, forming an extremely-wide-field ultra-high-cadence sky survey. The high-speed mode will be able to make cutouts to cover $\sim2\times10^7$ bright stars across the sky in each exposure.

\section{All-sky array key science cases}
\label{sec:science}
All-sky large telescope arrays will explore a unique parameter space, covering the entire sky with a cadence hundreds of times faster than any other deep sky survey. We here explore the key science projects unique to an all-sky array, with (where appropriate) specific numbers calculated for the Argus Optical Array based on the Argus-8 telescope.

\subsection{Multimessenger Astronomy}
Whether the events are detected by gravitational waves (GW), neutrino sources, or gamma ray detections, multi-messenger transients are typically rapidly detected over the whole sky, while optical follow-up must scramble to tile across the areas of interest. In constrast, the all-sky array will be the only system with "negative slew time", sensitive to events where the optical counterpart leads the other detections.

\textbf{Gravitational wave counterparts:} the all-sky array will be able to search for faint coincident counterparts to gravitational-wave events without requiring pointing, thus reaching the events as they occur. The system will be able to see the brighter ($\rm m_g \lesssim 20$) events over the entire event uncertainty region within one minute, and then reach depths comparable to current rapid response surveys such as DLT40 \citep[][]{Valenti_2017} and ZTF \citep[][]{Bellm-2019} within minutes, without suffering phase smearing for expected KNe evolution timescales. The array will thus be able obtain multicolor early-time lightcurves for KNe in its quarter-entire-sky field of view, constraining their central engines. 

For the lowest-redshift and brightest events, binary neutron stars (BNSs) will spend multiple minutes with detectable SNR in the band of the LIGO and Virgo detectors \citep{Sachdev_2020}. This may allow the detection of a forthcoming event tens of seconds before the companions collide and merge. These early-warning LIGO triggers \citep{Magee_2021} are largely infeasible to follow-up for most optical instruments due to the extreme slew time requirements and the large error region sizes. The all-sky array would be able to cover these events as they occur in a quarter of the entire sky, and potentially switch to high-speed exposures in all telescopes as the event occurs. This capability opens up the possibility of constraining precursor emission to GW events, or GRB prompt emission to study jet formation mechanisms.

\textbf{Increasing LIGO's event rate:} an all-sky large telescope array also has the potential to increase LIGO's event rate by searching for time-coincident optical counterparts to GW events which do not reach the criteria for formal detection; an optical detection increases the confidence in the gravitational wave event. Given the low number of binary neutron star mergers seen to date, each KNe is important to constrain the central engines, determine the heavy element production rates, and ultimately provide an independent measure of the Hubble constant. Depending on the model used for early-time evolution of the kilonovae \citep[][]{kn-model-lp98, kn-model-met10, kn-model-bk, kn-model-piran}, the Argus Array's optical/GW search will be capable of increasing the neutron-star-merger detection volume of LIGO/Virgo by a factor of ~2-8, with a corresponding increase in the event rate (Ackely et al. 2021, in prep). 

\textbf{Fast radio bursts:} recent surveys like CHIME have discovered dozens of fast radio bursts (FRBs), but their origins are still mysterious, despite a few events having been localized to host galaxies \citep[][]{chatterjee-2017, ravi-2019, bannister-2019, marcote-2020}. A variety of origins have been suggested, ranging from astroseismology on exotic stars to flares from highly-magnetic neutron stars \citep[][]{metzger-2019}. Many FRBs repeat, while others appear to be one-off events, or at least repeat on timescales beyond those probed by current surveys. This indicates that there may be multiple mechanisms for producing fast radio bursts. Optical counterparts would provide the leverage to nail down the origins of FRBs \citep[][]{Lorimer-2008, metzger-2019}, but no transient optical counterpart to an FRB has been discovered to date: the timescales preclude follow-up with conventional telescopes (except coincident-pointing for narrow field arrays, eg. \citet{bloemen-2016}). The array is able to monitor the entire Northern sky for FRB counterparts, including the entire CHIME FoV. This will guarantee dozens of simultaneous optical/radio observations of FRBs, at greater depths and faster cadences than have yet been probed, and addressing the FRBs which do not appear to repeat. The array will detect transient sources in realtime at minute-cadences, and in bright time could perform sub-second-cadence monitoring of all nearby galaxies within 20\% of the sky.

\subsection{General optical transients}
\textbf{Extremely-early-time supernovae:} unique information about the progenitors and explosion physics of transients is conveyed in their earliest stages \citep[e.g.][]{1999ApJ...510..379M, 2010ApJ...717..245K, 2014Natur.509..471G}. When the supernova shock breaks out of a progenitor star it releases a flood of high energy photons that ionize any material in the vicinity. So-called "flash spectroscopy" of the resulting recombination lines probes the chemical makeup of the progenitor wind, giving information about the state of the star just prior to its death \citep[][]{2014Natur.509..471G}. Furthermore, dense sampling of the light curve during shock breakout itself can reveal the structure of the progenitor \citep[e.g.][]{2018Natur.554..497B}. Current tiling sky surveys must balance sky coverage with timescale, and can detect very-early-time transients only in the fraction of time and sky devoted to ``deep-drilling'' survey fields. The Array's cadence will enable early detection and measurement of these events –- as well as the Type-Ia SNe used for supernova cosmology.
 
\textbf{Supernova cosmology:} early-time detection, well before event peak, is challenging for current sky surveys of the nearby events which are used to constrain dark energy, but vital for measurement of the light curve stretch factor. The depth and continuous monitoring of a large all-sky array has the chance to lead to the detection of Type-Ia supernovae (and other extragalactic transients) an order of magnitude faster than current untargeted surveys; furthermre, the light curve of each detection can be measured in two colors over the long term, automatically.

\subsection{Exoplanets}
\textbf{Microlensing:} an all-sky array will be able to search for distant icy worlds impossible to find via transits by monitoring $10^8$ stars across the sky for all microlensing events, and billions of fainter stars for high-magnification events. Almost all microlensing surveys (e.g. OGLE, MOA, KMTNet, UKIRT) have been performed with relatively large telescopes observing the large population of background stars towards the galactic plane. This strategy, also to be followed by the Roman Space Telescope, optimizes the number of detections for relatively narrow-field instruments, at the cost of the vast majority being too distant for follow-up of the detected planetary systems. However, occasional spectacular events around relatively nearby stars (e.g. \citet{gaudi-2008, Fukui-2019}) have demonstrated that a sufficiently large-sky-area survey can detect events from stars tens to even hundreds of times closer to us – enabling the detection of planets smaller than Earth in habitable-zone orbits, breaking mass and orbit degeneracies via proper motions, and even the subsequent direct-imaging follow-up of detected icy giant planets. Detecting nearby events requires an all-sky survey capable of monitoring the millions of stars required for a usable event rate, at the hour-level cadence required to detect small planets in these relatively fast events. For the Argus Array, we predict O(100) galactic events per year at minute cadence, and 500 at hour cadence \citep[][]{han-2008}. It is possible that additional events will be detectable in local-group galaxies \citep[][]{deJong-2008}. Detected events can be immediately checked for achromaticity via the system's two-color observing, and the array's cadence allows planet confirmation without other-telescope observations, greatly reducing the complexity of event confirmation.

\textbf{Transit and eclipse timing:} the array will also likely be capable of detecting tens of thousands of transiting planets, although most would likely be in the regime already well probed by TESS. The all-sky array is better suited to concentrate on the searches that truly take advantage of the array's depth and extremely-high cadence, such as minute-timescale compact-object transits and timing searches. The Argus Array could, for example, provide minute-cadence monitoring for $\rm9\times10^6$ eclipsing binaries with per-image SNRs of $>$30 and 100+ epochs across each eclipse. This dataset will enable a sensitive timing search for circumbinary planets, providing the most stringent measurements on planet formation in close binaries (e.g. \citet{cukier-2019}; the uncertainty in population implies that the Argus Array can detect between dozens and thousands of planets).

\textbf{Compact object transits:} at least 1/3 of WDs have metal contamination in their atmospheres suggesting that there could be disintegrating rocky bodies around them, offering perhaps the only feasible chance of determining the composition of extrasolar rocky bodies, including terrestrial planets. Objects transiting white dwarfs show deep, often minute-timescale eclipses due to the small WD diameter. Surveys thus require a very high observational duty cycle on each target to have a reasonable chance of detecting transits of material \citep[][]{vanderburg-2015}, unless there are large clouds of debris \citep[][]{vanderbosch-2019}. Only a few detections, after major surveys by GALEX, Kepler and TESS (e.g. \citet{vanSluijs-2018, rowan-2019}) suggest that these systems are exceptionally rare. The Argus Array will be able to search 125,000 white dwarfs for debris-disk-transits, an order-of-magnitude increase over current surveys. Assuming a 25\% debris-disk fraction and accounting for geometric factors and other constraints, we estimate that the Array will produce between 10 and 300 detections. Although statistics are difficult to estimate given the low incidence rate, it is also possible that an all-sky array will detect ultra-short-transits from giant and even rocky planets transiting white dwarfs (e.g. \citealt{Vanderburg2020}).

\subsection{Stellar Astronomy}
On tens-of-minutes timescales, the Argus Array goes deep enough to monitor at least 20$\times$ more stars than TESS's full-frame-images \citep[][]{2019PASP..131i4502F}. With the Array's fine pixel scale easily allowing 1\% photometry in all but the most crowded regions of the galactic plane, the Array will provide long-term, high-cadence data for essentially all Northern stellar astronomy targets. The resulting dataset will form a legacy survey enabling, among other projects, studies of stellar activity from all exoplanet hosts; precision timing measurements for all eclipsing binaries; long-term, high-cadence pulsation measures for all known white dwarfs and hot subdwarfs; outburst monitoring for all X-ray binaries; the deepest search for AM CVN binaries, which will be some of the strongest LISA GW sources; rotation rates confirming radial-velocity-detected planets; searches for star-planet interactions driving stellar flares; or mapping the galactic halo via deep searches for flares from halo M-dwarfs and pulsations of RR Lyr.

\subsection{Solar system objects}
\textbf{Rocky objects:} each minute the Argus Array is capable of measuring photometry for 40,000 main-belt asteroids, 400 near-Earth objects, 200 Trojans and around 30 Kuiper Belt objects; at least 10$\times$ more will be detectable in ratchet-scale coadds, and the numbers will increase still further for slower-moving objects easily detectable in multiple-hour coadds. 

For rapidly moving main-belt asteroids, where coadding images in unknown orbits is challenging, an all-sky array would require significant software work to detect objects deeper than the largest current surveys, such as Pan-STARRS. The population of unknown objects at the magnitudes accessible to an all-sky large telescope is likely to be small, and a Rubin-Observatory-like deep short-exposure will be obviously more effective at finding faint rapidly-moving objects.

An all-sky array would, instead, build an unprecedented set of $10^5-10^6$-epoch light curves for every target. This dataset will enable a comprehensive measurement of rotation rates, shape parameters, and thus internal strengths for minor planets across the Solar System, while also searching for asteroid moons via eclipses (potentially measuring asteroid masses). The array could provide the most comprehensive search for rare super fast rotating asteroids --- bodies $>$200m in diameter with periods $<$2 hours that require significant internal strength (e.g. \citealt{Thirouin_2018}). The array will also return high-cadence light curves for comets, searching for outbursts and changes in morphology. 

\textbf{Occultations:} by monitoring $10^8$ stars continuously at high cadence, the array will also be capable of detecting minute-timescale occultations by distant large ($10^3$ km-scale) objects in the Kuiper Belt or even Oort Cloud. Based on current population statistics, we estimate the Argus Array could detect a few such objects, and the Array's high-speed mode would potentially enable the detection of 100s of 100km-scale objects.

\textbf{Interstellar asteroids:} further afield, the recent discoveries of an interstellar asteroid \citep[][]{2017Natur.552..378M} and comet \citep[e.g.][]{2019ATel13100....1G} have opened a new field. Light-curve monitoring revealed `Oumuamua's possible highly-elongated shape. The object's anomalous acceleration \citep[][]{2018Natur.559..223M} is most easily explained by comet-style outbursts, but none were observed in the patchwork of coverage from major observatories.

Conventional deep-snapshot sky surveys are more suited to instantaneous detection of faint moving objects. Once they are found, an all-sky array can contribute by providing high-cadence light curves for all interstellar asteroids detected by current and planned Northern-hemisphere surveys, and will be able to follow-up Rubin-Observatory-detected objects as they get closer to the Earth. The array will be able to constrain orbits, measure rotation rates, enable shape measurement for complex tumbling objects, use the high-cadence multi-color light curves to constrain surface variations, and search for outbursts over long-term monitoring, for all objects across the sky each night without requiring individual targeting.

\section{Building and operating the Argus Optical Array}
\label{sec:design}

It is a simple concept to consider covering the sky using hundreds of wide-field telescopes. However, with thousands of individual components and moving parts, the array itself could become one of the most complex astronomical instruments yet devised. Actually building, operating and maintaining such a system is therefore a new class of astronomical instrumentation problem. In this section we discuss how to make a large array telescope work reliably and efficiently over a long-term survey.

\subsection{The challenges of operating a very large telescope array}

\begin{figure}
	\centering
		\includegraphics[width=\columnwidth]{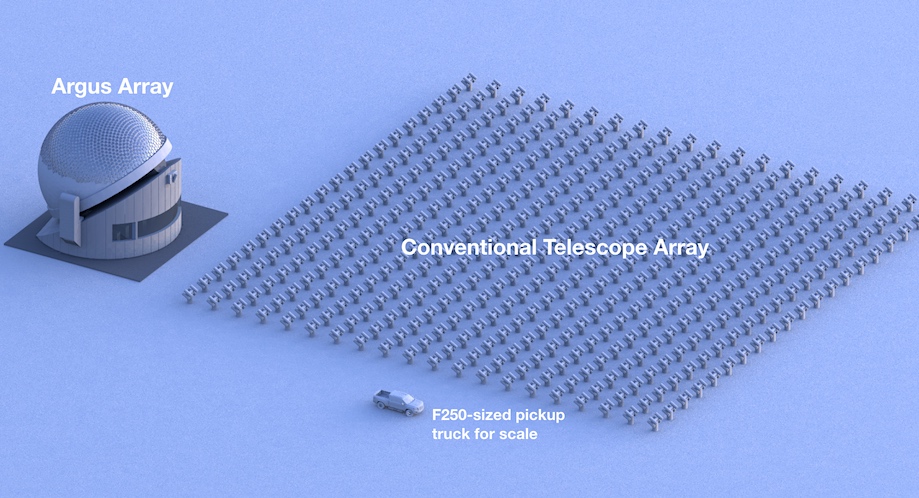}
	\caption{A comparison of the Argus Optical Array hardware to a conventional telescope array of the same aperture. Each array hosts an identical number of telescopes (here the Celestron RASA-11); we assume each conventional mount hosts four telescopes. A large pickup truck is shown for scale. For an all-sky survey strategy, the Argus Array is a physically much smaller system which keeps all telescopes within a stable, clean environment, and thus greatly reduces construction and operations costs as well as maintenance downtime. This comes at the cost of flexibility in survey design.}
    \label{fig:telescope_design_comp}
\end{figure}

Each of the hundreds of telescopes in the array must track the sky, focus, cool its camera, and provide consistent image quality. The conventional telescope array concept, with small groups of telescopes on individual mounts, becomes complex at this scale (Figure \ref{fig:telescope_design_comp}). Hundreds of tracking drives must maintain arcsecond-level performance, focusers on each telescope must maintain image quality though large outside temperature changes, the telescopes must maintain consistent point spread functions across their fields, the cameras must keep operational over long periods in outside conditions, and a total of thousands of optics exposed to dust, dirt and thermal cycles must be kept clean and well-aligned.

If each telescope and tracking drive requires a total of only a few hours of cleaning and maintenance each year then multiple full-time operators will be required just to perform that routine maintenance. If and when more complex maintenance such as routine recollimation is needed, years of work would be required. Since the extremely-long maintenance timescales preclude a conventional engineering-time shutdown, a large telescope array will necessarily observe with a constantly-shifting set of telescopes with different optics cleanlinesses, image qualities, and tracking performances.

\subsection{The Argus hardware design}
We have designed the Argus Optical Array to reduce the operations requirements of a large telescope array to  at or below the costs of a single large survey telescope. The Argus concept maintains the telescopes within a single sealed, thermally-controlled environment with a single moving part used during the night for tracking. The telescope array, rather than consisting of thousands of individual moving parts, therefore acts as a single ``solid-state'' unchanging system in a stable lab-like environment. The costs and complexities associated with making the individual telescopes cope with outdoor conditions are eliminated, and the sealed and temperature-controlled telescope environment ensures that minimal ongoing maintenance of the individual components will be required.

\subsubsection{Telescope dome structure}
The hemisphere internals, telescopes and cameras, will be kept at an internal temperature of $10\pm0.5^{\circ} \rm{C}$, with a stability level chosen to eliminate the need to refocus the telescopes. Internal partitions will allow the maintenance of temperature within the focus stability limits without dealing with the complexities of large-scale convection currents within the dome. This arrangement keeps the optical systems stable and clean over long-term operation, thus reducing risk and costs (the exact internal temperature will be optimized for the chosen site's average temperatures, and potentially adjusted seasonally). Each telescope will observe through an antireflection-coated window with a gasket seal, with an air-gapped internal window preventing the formation of condensation on the telescope glass on the warmest nights. The dome structure will be aerodynamic compared to an open telescope structure, and the dome will be isolated from the telescopes, minimizing wind shake. Insulation and careful selection of the internal temperature will ensure the surface of the dome is close to ambient temperature, mitigating telescope-induced seeing. 

With a sealed structure, there is no need for an external dome to protect the array. Each individual telescope will instead use a simple and very reliable shutter to stop sunlight from reaching the optics during the day. This arguably adds one moving part per telescope, but these are extremely-low-precision systems that are used only twice per night, and do not affect the telescope imaging during the night. A larger-area backup system will be available to protect the array in the event of a shutter failure.

Because the telescope structure itself is weather-sealed, and the telescope optics are exposed only to filtered air in a controlled environment, we anticipate that regular optical cleaning will only be necessary for the external windows. These will be engineered for weatherproofing, and can thus be cleaned with aggressive methods: the personnel and equipment required to clean all optics exposed to unfiltered air will be a single person with a pressure washer, cleaning the telescope regularly. Fixing optics exposed to pollen, wildfire ash, and other antireflection-coating-destroying environmental effects will require only replacement or recoating of flat, mass-produced glass (the system can be kept covered by the emergency sunshade blanket in the worst events).

\subsubsection{Array arrangement and dome size}
The optimal packing of rectangular telescope fields of view across the spherical sky has a very wide range of possible solutions, depending on the survey parameters to be optimized. Fortunately, we can reduce the parameter space of the problem. We require that the telescopes form stripes of continuous declination, which ensures that fields maintain roughly the same positioning in subsequent cameras from one ratchet to the next. We also require a small overlap to allow photometric solutions between fields. 

Next, we must ensure that the telescopes do not physically overlap within the dome.  For all telescope placements we make sure the angle between the telescope sky pointing positions is larger than $\arccos(1-[\frac{d_t}{2r\cos(d)}]^2)$, where r is the radius of the sphere formed by the backs of the telescopes inside the dome, $d_t$ is the diameter of the telescopes, and d is the declination of the telescope under consideration. Over most of the sky, this condition is satisfied with a simple overlapping grid of telescopes. Near the pole, we require manual adjustment of the array positions to avoid physical telescope overlap.  For declinations greater than 80 degrees the rows are placed closer together, separated by 60\%-85\% of the field of view, in order to achieve continuous sky coverage (Figure \ref{fig:sky_coverage}).

Together, we find that a sky coverage of 7,916 sq. deg. with an average of 2.64\% overlap can be achieved using 900 Argus-8 telescopes. This requires a 45-foot diameter dome, although with support and ancillary structures between the telescopes this will be extended to 50-55 feet diameter. We are currently evaluating the design for the overall dome structure; our current concept is a geodesic dome with individual camera mounts in the faces of the dome. This structure has the advantage of strength, modularity and ease of construction; there are many companies undertaking geodesic-dome-engineering for the home-building and commercial construction industries. 
We are also considering an alternative approach of spreading the telescopes over a small number of smaller domes. The Argus prototypes will allow a measurement of the engineering tradeoffs associated with this design.

\subsubsection{Thermal control and seeing}
The sheer number of cameras required for an array the size of Argus has the potential to become a problem regardless of the array design: during normal operation, the thermoelectric cooling required for 900 IMX455-based cameras will produce up to 60kW of heat. In a conventional telescope array design, this heating would produce dozens of columns of hot air with a potential to cause seeing issues. By maintaining all cameras within the Argus dome, the large majority of the cooling power can be supplied by an efficient central A/C system with remote condenser coils, relying on the camera coolers only for precision temperature stability, and eliminating the concern of heat-induced seeing. Internal insulation will ensure the surface of the dome is close to ambient temperature, mitigating telescope-induced seeing and condensation. 

The Cassegrain-focus of the Planewave Instruments Argus-8 telescopes greatly facilitates this design, by placing the cameras behind the telescopes where thermal load can be dumped into specialized systems. The prime-focus camera position of the RASA telescopes places the camera heating and fan in front of the telescope primary, and next to the window of the system in the Argus design, necessitating a much more complex thermal control system. The under-construction Argus prototypes are scaled to validate the performance of the thermal control systems under a variety of external conditions.

\subsubsection{Tracking} Unlike standard telescope mounts, Argus places the system's tracking drive at the edge of the hemisphere, the entire dome being mounted onto an axle with a high-capacity bearing pointed at the celestial pole. This design takes advantage of the resulting large lever-arm, a linear actuator drive, and spring-tensioning to track the hemisphere smoothly and with strong wind-shake resistance.  We plan to track the system for 9 minutes, then ratchet back for 48 seconds, then return to tracking. In this way, each field will swap from one camera to the next one in its declination stripe every 9.8 minutes. The camera arrangement will be optimized to keep stars in the same regions of sequential cameras, reducing PSF variations ratchet-to-ratchet; multiple-color surveys will be performed by alternating color filters on sequential cameras. To allow consistent image subtraction and reduce the variation of point spread functions for image coaddition, the system observing schedule will be fixed so that each tracking position is selected from a set of fixed fields.

\subsection{Data analysis pipeline}
\label{sec:pipeline_design}
In the same way that careful hardware design is required to control operations costs to reasonable levels, careful attention to software scope is required to avoid losing those savings in pipeline complexity. The Argus Array will produce 54.9 GPix of raw data every minute, over 100 PB of raw data over the course of the project. Fully recording this enormous data rate would conventionally require a hugely-expensive data analysis architecture which would make for an enormous software project with a budget far beyond to the cost of the telescope hardware. We argue that, with careful attention to project scope, it is possible to produce a scientifically-productive array at far lower (and thus much more feasible) costs. 

\subsubsection{Scope control}
We would suggest that the most important consideration is to trim the software and data storage requirements to the minimum, building extra capabilities later if required. The Argus science cases we have identified do not require tightly-controlled PSFs across the field, better-than-few-\%-precision photometry, supporting more than one survey strategy, supporting many cadences, supporting many passbands, long-term storage imaging data at the full incoming rate, storing high-cadence data at all (beyond selected targets and some transients), developing a new data analysis pipeline from scratch, making all data available in a full querying interface, or supporting data products beyond the basics of light curves, images and transient alerts. 

The array will likely be able to perform other surveys such as transit detection, with additional software work. Eliminating, however, high-precision, high-performance and complex query interface requirements from the initial pipeline design would allow an all-sky array's most unique science to be produced more rapidly. Importantly, it would also reduce the project initial software cost to a level much more commensurate with the off-the-shelf telescope hardware, compared to attempting to build a major-facility-class pipeline capable of addressing every possible science program from the beginning. 

\subsubsection{The Argus hierarchical pipeline design}

\begin{figure}
	\centering
		\includegraphics[width=\columnwidth]{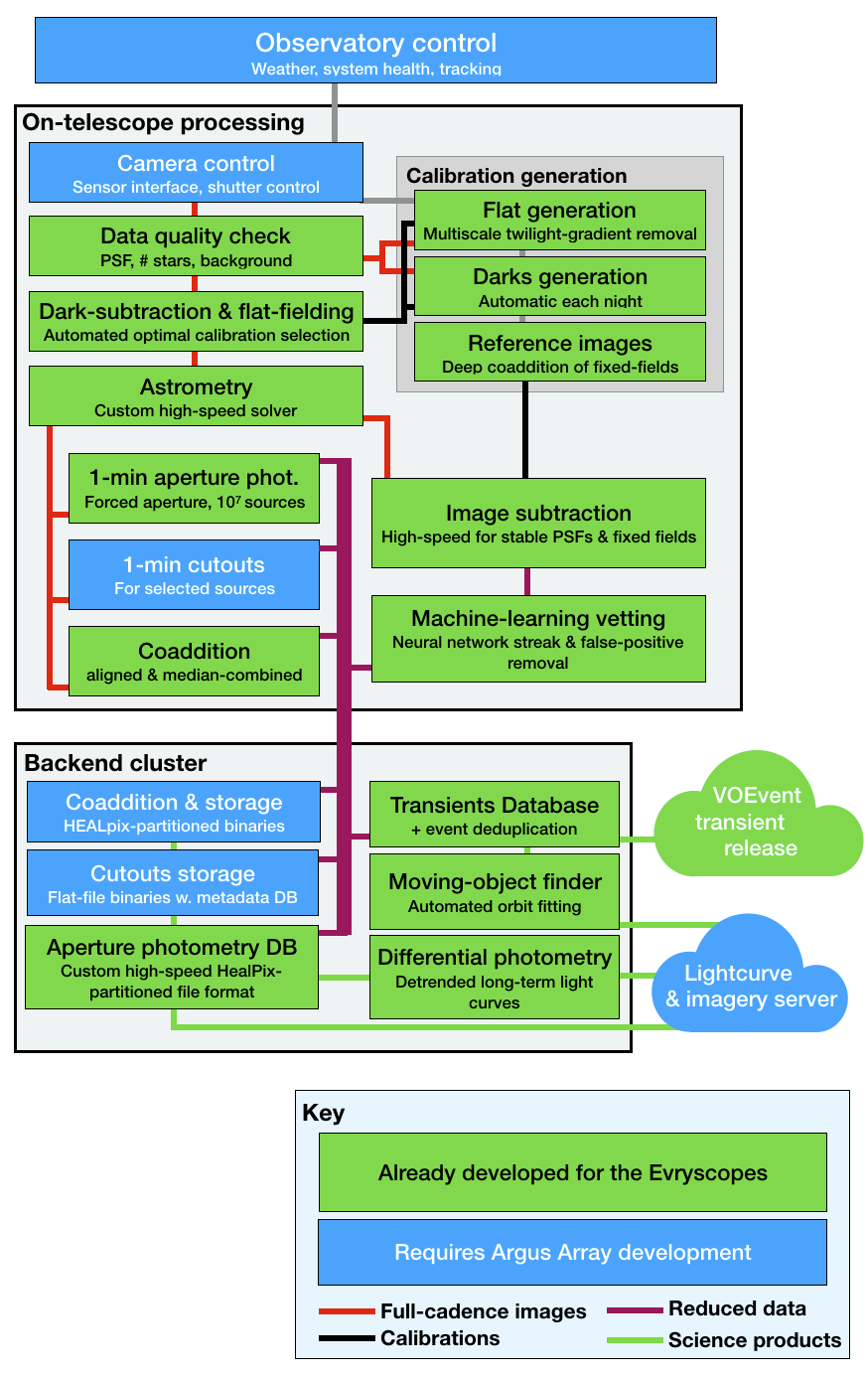}
		
	\caption{A block diagram of the hierarchical pipeline design which makes the Argus Optical Array data analysis practical. A high-speed subsystem processes incoming data and only sends a small, lower-cadence subset to the main analysis cluster for recording. 9-minute-cadence summed images are stored, along with higher-cadence thumbnails of areas with changing sources. The high-speed burst mode bypasses most of these steps by either storing minutes of high-speed images directly for later, slower, analysis, or (for longer-term observations) selecting cutout positions at the start of a ratchet, and directly storing high-speed data for those regions only.}
    \label{fig:pipeline_design}
\end{figure}        

The Argus hierarchical data processing system (Argus-HDPS) is designed to bring the scale of the all-sky-array data-analysis challenge to a level already demonstrated by the current Evryscope systems, by only recording the most scientifically-important data from the individual telescopes. Like high-data-rate particle physics experiments, or large radio telescope arrays, the Argus-HDPS (Figure \ref{fig:pipeline_design}) processes all data with an initial computing layer that a) stores all data in the short-term and b) streams only the scientifically-interesting parts to long-term storage. Minute-cadence data is stored for tens of millions of targets selected by the science team; real-time minute-cadence image subtraction and transient detection will be performed; and the full-frame images are coadded and stored at a one-ratchet cadence. A 1000-core local cluster, available off-the-shelf from multiple manufacturers, will perform the longer-than-ratchet coaddition, store the data products and make them available to the astronomical community. A local high-speed cache will support burst-mode high-time-resolution telescope exposures.

A key engineering requirement of the data analysis system is that each telescope's analysis system is responsible only for processing data from its own detector at the highest-level cadence, without using any data from any other telescopes. This guarantees that the data analysis, up to the slower coadding of multiple ratchets, is an embarrassingly-parallel algorithm; adding more telescopes to the system thus only linearly increases the total system load, making the analysis scheme feasible for any reasonable number of telescopes.

In the Argus Array, each telescope is planned to stream data over a fiber link to a server in a nearby cluster, running a modified version of the Evryscope pipelines. This system is capable of performing dark-subtraction, flat-fielding and astrometry of incoming data in real-time in a single core (\citealt{2020ApJ...903L..27C}; a lab-demonstration system has performed this for Argus images even on low-power Raspberry-Pi CPUs). First, all incoming data will be transferred to a short term (two-week-timescale) data store as contingency for unusual and exciting events. Second, in realtime, postage stamps around science-team-selected targets will be extracted, and Evryscope-pipeline-validated image subtraction will be performed on the images to detect and store stamps for new targets. Finally, the images will be co-added to a one-ratchet cadence, stored as full frame images, and passed to the backend cluster for longer-timescale coaddition. The Argus Array will be backed by an integrated storage/analysis system to reduce the incoming images into multi-scale, time-series data products. Data from the per-telescope compute units will be streamed to the local data warehouse with a petabyte-class enterprise storage system; the compute cluster will perform deep coaddition and transient discovery across multiple cameras. Using the existing Evryscope infrastructure, the ratchet-cadence full-frame images will be split into smaller subsections, and indexed by position on a HEALPix grid, with metadata stored in a quick-access database system. High-cadence postage stamps will be stored in a similar system.

When operating in the high-speed burst mode, the Argus-HDPS will perform a subset of the standard-cadence analysis steps, optimized for the sub-second cadence regime. New sources will be detected using a simplified image subtraction algorithm designed for readout-dominated exposures with negligible tracking minimal or PSF variability between images. Transient sources will be deduplicated against a science-team selected target catalog and reference sources, and small cutout stamps extracted for later analysis.

\subsubsection{Data products}
Each image produced by an Argus telescope will contain 10,000 -- 30,000 science targets, totalling up to $10^8$ targets in each whole-array exposure.  To reduce the data rate to feasible levels, high-cadence light curves can be automatically generated for targets selected by the Argus science team, based on the small cutout stamps generated by the first layer of the pipeline; this method of data-load-reduction has long experience from space missions such as TESS \citep{Ricker_2015}. Aperture photometry will be performed within the cutouts, with the local background calculated by interpolation across the full images adjusted by the priors measured within each cutout. Like the Evryscope pipelines \citep{ratzloff-evryscope}, differential photometry is most easily performed using a grid of carefully-selected reference stars stored along with the science targets. 

Argus will use the existing Evryscope pipelines for robust image coaddition for faint transient discovery. The per-telescope servers will coadd images each ratchet, performing simple coadditions enabled by the short timescale, stable pointing and smaller differential atmospheric refraction terms from the short ratchets compared to Evryscope. The backend cluster will combine multiple camera's individual coadding results, taking advantage of the reduced data rates to use more sophisticated (and slower) optimal algorithms, including modern, statistically-robust tools \citep[][]{1999AJ....117...68S, Zackay-2015a, Zackay-2015b}.

\subsubsection{Satellite constellation effects}
Starlink and other similar satellite constellations will increase the number of satellite trails by orders of magnitude \citep[][]{McDowell_2020}. The  exceptionally high cadence of the all-sky array mitigates these concerns greatly compared to a conventional tiling sky survey. With 60 images per hour of every part of the sky, trail detection and outlier rejection will enable the efficient removal of satellite-affected image regions. Transient detection will be confirmed by requiring a second image, taken within a minute. For high priority targets (e.g., known planet hosts with high-amplitude flaring, or galaxies within the stamp of a gravitational-wave detection), alerts and followup resources can be triggered based on a location prior, either in addition to or in place of the confirmation image.

\subsection{Overall project budget}
We estimate that the optics and cameras for an Argus-8-based all-sky array design can be purchased, at mid-2021 list prices, for a total of \$6.1M (this will likely be reduced by bulk purchasing). The dome building, including subcontracts for the required telescope mounting systems, cabling, power supplies, and other accessories, totals approximately \$3M. The off-the-shelf data storage and analysis clusters detailed above total \$2.1M, for a total hardware cost of \$11.2M. 

Three-year site costs (power, internet, etc.) total \$0.7M. We estimate that, leveraging existing software pipelines and the already-funded engineering work, the total personnel costs for the construction and commissioning of the Argus Array and its pipelines will be approximately \$5M. We estimate that the full array can therefore be constructed and operated for an initial survey within a total budget of approximately \$17M. The upgraded 11-inch aperture array, capable of almost one magnitude greater coadded depths, would total around \$25M.

\section{Prototypes and Conclusions}
\label{sec:conclusions}
\subsection{All-sky-array Prototypes}
The Argus Array is currently in the prototype stage. Funded by an NSF CAREER grant, an NSF Mid-Scale Innovations Program development investment, and a grant from Schmidt Futures, the prototypes are designed to demonstrate the  concept at the medium (dozens of telescopes) scale, including linear scaling of all hardware and software operations to the full system. This staged test program includes:

\begin{enumerate}
    \item {\textbf{EvryArgus:} An individual Argus telescope tested in North Carolina in 2019, and mounted on top of the Northern Evryscope at Mount Laguna Observatory in California in early 2020. This system was shut down by the COVID-19 pandemic.}
    \item {\textbf{milliArgus:} A 16-telescope testbed system contained in an 8-foot thermally-controlled dome. milliArgus is designed to validate the long-term performance of the Argus tracking drive design, telescopes, cameras and thermal-control systems and rapidly retire risks associated with the development of the Argus technology, as well as produce data for pipeline development and testing. milliArgus is currently under construction and will be deployed to the Pisgah Astronomical Research Institute dark-sky site in North Carolina in mid-2021.}
    \item {\textbf{Argus Pathfinder:} The Argus Pathfinder will be a larger-scale system based in a 20ft-scale dome containing 48 Argus telescopes and cameras. The Pathfinder, with a stripe of telescopes covering all declinations simultaneously, will observe each part of the accessible sky for 9 minutes per night (more near the poles). Although its limited RA coverage will prevent it approaching the coadded depth of the full Argus Array, it will allow the first extremely-high-cadence, wide-field science to be performed across the whole sky each night. The Pathfinder will also pioneer the extremely-high speed observing modes, and the associated pipelines and computing systems. The Pathfinder array, currently being designed, is fully funded by the National Science Foundation and Schmidt Futures, and is planned for on-sky operation at Mount Laguna Observatory in 2022.} 
\end{enumerate}

Subsequent papers will detail the on-sky performance of the Argus test systems. 

\subsection{Conclusions}
The new mass-produced, low-cost, wide-field telescopes have the potential to enable entirely new types of large telescopes: giant arrays capable of imaging the sky with many-meter-class collecting areas, at costs more than an order of magnitude lower than monolithic single-mirror telescopes. Beyond just large and low-cost collecting area, these arrays open the possibility of new types of surveys impossible to achieve with a standard telescope, including 1) uniquely flexible surveys dynamically trading off sky coverage for depth; or 2) enormous-field-of-view surveys which image the entire accessible sky simultaneously, achieving depth with night-long exposure times. Together, these systems will be able to explore a host of ultra-fast, rare phenomena which cannot be surveyed with conventional telescopes.

The large-array telescope opportunity, however, comes with a risk: these arrays would be, arguably, the most complex astronomical instruments yet designed, with thousands of moving parts and optics exposed to dirt and thermal cycles. Very careful attention to the prosaic maintenance and operations costs is required to actually achieve the potential science return from large arrays, without becoming lost in a mire of constantly-changing broken telescopes and disappointing image quality. 

In this paper, we presented one possible solution to these challenges, where all the optics are kept sealed in a thermally-controlled environment and the number of moving parts used during the night is reduced to a single tracking drive. We are currently building prototypes to demonstrate this design and start the first high-cadence all-sky observations with the survey telescopes and their high-speed detectors. The 2022 completion of the Argus Pathfinder will bring the Argus project to the Preliminary Design Review stage, enable the first all-sky explorations of Argus-timescale science, and pave the way to the construction of the full Array.

\samepage{
\section*{Acknowledgements}\label{acknowledge}
We thank Eran Ofek, Sagi Ben-Ami, Enrico Segre and Angelle Tanner for helpful conversations. This paper was supported by NSF MSIP (AST-2034381) and CAREER (AST-1555175) grants, and by the generosity of Eric and Wendy Schmidt by recommendation of the Schmidt Futures program. This research, and the construction of the Argus prototypes, is undertaken with the collaboration of the Be A Maker (BeAM) network of makerspaces at UNC Chapel Hill and the UNC BeAM Design Center.
}

\appendix

\section{Argus hardware trade study}

The Argus Array system design concept depends on the commercial availability of wide-field telescopes and large image sensors with few-micron pixels and high sensitivity. The requirements for the individual imaging systems include both science-driven requirements to maximize survey depth, and practical requirements for ease of construction. Collecting area, resolution, and FOV must be optimized for survey depth and speed, while simultaneously minimizing cost, weight, and the total number of telescopes.

\subsection{Telescopes}
Celestron's Rowe-Ackermann Schmidt Astrograph (RASA) line of telescopes \citep{rasa_whitepaper} includes 8, 11, and 14-inch versions. Additionally, a custom telescope (Argus-8) is currently under development in collaboration with Planewave Instruments, which will produce a finer pixel scale than the RASA-8 with improved optical quality. Table~\ref{tab:cots_optics}  details the relevant parameters for the three Celestron RASA telescopes and the Planewave Argus-8.  $N_{sky}$ is the number of telescopes needed to tile an 8,000 sq. degree FOV assuming $\sim$3\% field overlap with 35 mm-format image sensors. Plate scales are calculated based on the 3.76 $\mu$m pixels of the Sony Exmor R image sensors described below. $D_{equiv}$ is the diameter of a single unobstructed circular mirror with a collecting area equal to an array of $N_{sky}$ telescopes. FOV and $N_{sky}$ for the RASA-8 exclude the ~18\% of a 35 mm format sensor that falls outside of the nominal image circle, and the values for the RASA-11 represent the second iteration of that design, introduced in 2020 with an improved focusing system. The RASA-14 was excluded from Argus Array testing due to its higher price point (\$141K per m$^2$ of aperture, relative to the sub-\$60K per m$^2$ of the RASA-8 and RASA-11), prohibitive $N_{sky}$, large weight, and large image circle requiring expensive large-format image sensors for optimal use.

\subsection{Sensors}
Due to the wide FOV and small image circles created by the RASA systems, an image sensor with
few-micron scale pixels is required to sample the sky at arcsecond precision. Sony
Semiconductor manufactures a variety of back-side
illuminated (BSI) CMOS image sensors with a pixels comparable to the optical spot size of the RASA optics
as part of their Exmor R and Starvis product lines. These sensors utilize stacked correlated-double
sampling both before and after per-pixel analog-digital conversion to achieve extremely low
readout noise while maintaining peak quantum efficiency of 90\% on most models. 

For Argus Array Prototype testing, we considered four 3.76$\mu$m-pitch Sony BSI sensors: the APS-C
format IMX571 (26 MP), the 35 mm IMX455 (61.2 MP), and the medium format IMX461 (102 MP) and IMX411
(151 MP).  All are available packaged as cooled astronomical cameras from
QHYCCD (all), Atik Cameras (IMX455), and ZWO (IMX571, IMX492, and IMX455).  All
3.76 $\mu$m sensors have low (1.7 e-) read noise in their high-gain configurations, dark current of
less than 0.01 e- per second at -5 C, and native 16-bit ADCs; they differ primarily in pixel count
and supported readout modes. Cameras based on the IMX461 and IMX411 are available from QHYCCD, but
with steep price scaling due to the physically larger sensors. 

\begin{deluxetable*}{lcccccccccccc}[!b]
    \tablecaption{\label{tab:cots_optics}Telescope parameters for an Argus-like system built around the
    three different Rowe-Ackermann Schmidt Astrograph (RASA) systems available from Celestron, plus
    our under-development custom mass-produced 8-inch telescope with improved optical performance and pixel scale. Pixel scale, and FOV are calculated for a Sony IMX455-based camera.}
    \tabletypesize{\footnotesize}
    \startdata
    \vspace{0.25cm}\\
    \bf{Telescope} & \bf{Dia.}     & \bf{Weight} &  \bf{f/}  &    \bf{FOV} &  \bf{Field Dia.} &  \bf{Spot Dia.} &    \bf{Pix. Scale}  & \bf{$D_{equiv}$} & \bf{$N_{sky}$}  &     \bf{Avg. 5-$\sigma$}   \\
                   &  (cm)         &  (kg)       &           & (deg$^2$)   &  (mm)            & ($\mu${m})      & (\arcsec per pixel) &    (m)           &                 &                  60 s (g'/r')   \\
    RASA-8         &  20.3         &   7.7        &  2.0      &   14.41     &  32              &  $< 4.55$       &        1.93         &   3.84           &    565          &          19.5/18.7    \\
    Argus-8        &  20.3         &   \nodata       &  2.8      &   9.04      &  43.3            &  $< 4.50$        &        1.38         &   4.83           &    900                  &         19.8/19.0   \\
    RASA-11      &  27.9         &   19.5        &  2.2      &   7.37      &  52              &  $< 4.50$       &        1.25         &   8.39           &    1103             &        20.4/19.5    \\
    RASA-14        &  35.6         &   34.0        &  2.2      &   4.54      &  70              &  $< 6.3 $       &        0.98         &   13.38          &    1791            &         20.9/20.0    \\
    \hline
    \enddata
\end{deluxetable*}

\bibliography{refs}{}
\bibliographystyle{aasjournal}

\end{document}